\documentclass{jfm}
\usepackage{color}
\begin{document}

\newtheorem{lemma}{Lemma}
\newtheorem{corollary}{Corollary}

\shorttitle{Noisy transitional flows} %for header on odd pages
\shortauthor{C. Lissandrello, L. Li, K. L. Ekinci, and V. Yakhot} %for header on even pages

\title{Noisy transitional flows in imperfect channels}

\author
 {
 C. Lissandrello\aff{1},
 L. Li\aff{1},
  K. L. Ekinci\aff{1}\corresp{\email{ekinci@bu.edu}}
  \and
  V. Yakhot\aff{1}\corresp{\email{vy@bu.edu}}
  }

\affiliation
{\aff{1} Department of Mechanical Engineering, Division of Materials Science and Engineering, and the Photonics Center, Boston University, Boston, Massachusetts 02215, USA}

\maketitle

\begin{abstract}
Here, we study  noisy transitional flows in imperfect millimeter-scale channels. For probing the flows, we use  microcantilever sensors embedded in the channel walls. We perform  experiments in two nominally identical channels. The different set of imperfections in the two channels result in two random flows in which high-order moments of near-wall fluctuations differ by orders of magnitude. Surprisingly however,  the lowest order statistics in both cases appear qualitatively similar and can be described by a proposed noisy Landau equation for a slow mode. The noise, regardless of its origin, regularizes the Landau singularity of the relaxation time and makes transitions driven by different noise sources appear similar.
\end{abstract}

\section{Introduction}
More than a century ago, Osborne Reynolds investigated transition to turbulence in a glass pipe in which he injected a filament of dye  at the inlet~\citep{Reynolds_1883}. Reynolds noticed that the characteristics of the dye filament and hence the entire flow field depended on the dimensionless flow rate or the Reynolds number,  ${\rm Re}=UD/\nu$ (here, $U$ is the mean flow velocity, $D$  the pipe diameter, and $\nu$ the fluid kinematic viscosity).  When $\rm Re$ was below a critical value,  $\rm Re\ll Re_{cr}$, the dye propagated downstream as a thin filament without breaking up, indicating a laminar flow in the pipe. At $\rm Re\geq Re_{cr}$,  this pattern changed dramatically:  waves  appeared in the vicinity of the inlet, sometimes leading to substantial flow  randomization downstream. With increasing $\rm Re$, the fraction of the tube length occupied by these  waves increased,  and at $\rm Re \gg Re_{cr}$, the entire flow became turbulent.

Reynolds, however, was unable to determine a \textit{unique} value for $\rm Re_{cr}$.  He noticed that the value of $\rm Re_{cr}$ was sensitive to various  imperfections, most notably the  geometry of the inlet. If the inlet was sharp,  inlet perturbations appeared in the form of shedded vortices that caused transition to turbulence at large Reynolds numbers. These perturbations, however, rapidly decayed downstream, if Reynolds number was not too large.  By carefully eliminating these, Reynolds was able to delay the transition up to $\rm Re\approx 12,800$~\citep{Reynolds_1883}.  Others following Reynolds  were able to sustain laminar flows in pipes even for Reynolds numbers as high as $100,000$~\citep{Pfenninger_1961}.   Relatively recently, the effects of initial (inlet) perturbations on $\rm Re_{cr}$ were quantified by introducing well-controlled disturbances (jets) injected through  holes in the vicinity of the inlet~\citep{Darbyshire_1995}.  Consistent with Reynolds' observations, the flow became turbulent at smaller and smaller Reynolds numbers  as the ratio of the disturbance velocity  to the mean flow velocity was increased. (For a comprehensive review, see \citet{Yaglom}.)

The above-described ``decay or amplification'' of waves (or perturbations) forms the basis of Landau's phenomenological theory of transition \citep{Landau_Lifshitz}. It  must be stressed that, based on his general approach which revolutionized  the theory  of  critical phenomena, Landau was not interested in the notoriously difficult  and non-universal problem of deriving the value of the critical (transitional) Reynolds number $\rm Re_{cr}$. Instead, he assumed the existence  of a transitional (marginal) velocity distribution ${\bf{\bar u}}$ at   $\rm Re=Re_{cr}$, and attempted an investigation of the general and universal flow properties in the vicinity of $\rm Re_{cr}$ in terms of a small perturbation velocity ${\bf v}$. Then, under the small (infinitesimal) perturbation, the total velocity becomes ${\bf u}={\bf{\bar u}} + {\bf{v}}$ and the total pressure becomes $p=p_0+p_1$. In general,  the field ${\bf{\bar u}}$ can be time-dependent but, following Landau, we assume the transitional pattern to be steady. The Navier-Stokes equations  can then be written as
\begin{equation}
{\bf \bar u\cdot \nabla \bar u}=-{\frac{{\bf \nabla} p_0} {\rho}} +\nu\nabla^{2}{\bf \bar u},\hspace{1cm} {\bf \nabla\cdot \bar u}=0;
\label{eq:NS_original}
\end{equation}
\begin{equation}
\frac{\partial {{\bf v}}}{\partial t}+{\bf \bar u \cdot \nabla \bf v} +{\bf v \cdot \nabla \bar u} +{\bf v \cdot \nabla v}=-{\frac{{\bf \nabla} p_1} {\rho}}+\nu\nabla^{2}{\bf v},\hspace{0.25 cm}{\bf \nabla\cdot  v}=0.
\label{eq:NS_perturbation}
\end{equation}
Both $\bf \bar u$ and $\bf v$ vanish on solid walls. Neglecting the ${\cal O}(v^{2})$ contribution to Eq.~(\ref{eq:NS_perturbation})  results in linearized  Navier-Stokes equations to be used for investigating instabilities in fluid flows. Now, the task is to find a solution to Eqs.~(\ref{eq:NS_original}-\ref{eq:NS_perturbation}) describing the time evolution of an initially  ($t=0$) infinitesimally  small perturbation ${\bf v}$. In this case, the ${\cal O} (v^{2})$ contribution to Eq.~(\ref{eq:NS_perturbation})  is neglected. For the perturbation, Landau assumed the form ${\bf v}({\bf r},t) = A(t){\bf f}({\bf r})$, where $A(t)= {\rm constant} \times  e^{-i\Omega t}$ is the slowly-varying amplitude with a complex eigenfrequency, $\Omega=\omega+i\gamma$, and ${\bf f}({\bf r})$ describes the spatial dependence. If the near-wall effects are not  too important  and in the frame of reference moving with mean flow,  one can write the differential equation
\begin{equation}
\frac{d|A|^{2}} {dt} = 2{\gamma _1}|A{|^2} - \alpha |A{|^4},
\label{eq:landau_nopert}
\end{equation}
which is to be solved subject to initial perturbation $A(0)=A_{0}$. The solution to this equation is
\begin{equation}
{\frac{|A(t){|^2}}{|{A_0}{|^2}}} = {\frac{{e^{2{\gamma _1}t}}} {1 + |{A_0}{|^2}{\frac{\alpha} {2{\gamma _1}}}({e^{2{\gamma _1}t}} - 1)}}.
\end{equation}
Indeed, setting $\gamma=c\rm(Re-Re_{cr})$, one can reproduce the observed ``decay or amplification'' of perturbations. When $\rm Re-R_{cr}<0$, $A(t)\rightarrow 0$ in the limit $t \rightarrow \infty$ ($t \gg1/\gamma$);  otherwise,  the amplitude grows, saturating at $|A(\infty)|\propto \sqrt{\rm Re-Re_{cr}}$.  The  theory  suggests that information about the  initial conditions disappears on a time scale $\tau = {\frac{1}{2 |\gamma_1|}} \propto \frac{1}{\rm |Re-Re_{cr}|}$. This behavior is similar to the ``critical slowing down''  in the proximity of  a critical point in phase transitions.  In pipe flows, it  has important and interesting consequences. A perturbation occurring at position $l$  and being advected by a mean flow of velocity $U$  stays in the pipe for a time interval $t_0 \approx (L-l)/U$, where $L$ is the pipe length. Therefore,  to observe the decay of  a perturbation generated in the  {\it bulk} or its growth  toward a final turbulent state, one needs $t_0 \gg \tau$, requiring unreasonably long pipes around $ \rm Re_{cr}$. The divergent relaxation time $\tau\propto |{\rm Re-R_{cr}}|^{-\zeta}$  with $\zeta\approx  0.56$ was also recently obtained  in numerical simulations of transition in  force-driven Navier-Stokes equations by \citet{Mccomb_2014}.

Landau's theory of transition,  though insightful, is better applicable  to wakes in flows past bluff bodies~\citep{Sreenivasan_1989} and in convection, i.e., in situations where  wall effects and viscosity do  not dominate.   It is  well-justified in pipe flows when the characteristic size of  a perturbation is ${\cal O}(D)$   and wall effects are unimportant. Efforts to describe transition in pipes using the Landau theory, most notably by Stuart and others~\citep{Stuart_1971}, focused on obtaining the magnitudes of the coefficients $\gamma_{1}$ and $\alpha$ and testing their possible universality. This universality remains elusive, suggesting that wall effects must be important in transition. Numerical simulations and experimental data show that, at least in the range $2300\leq { \rm Re} \leq 10^{5}$, powerful bursts generated by unstable boundary layers are mainly responsible for turbulence production in the bulk.

The majority of workers studying transition to turbulence in pipes have been interested  in  the flow  response to perturbations in otherwise perfect pipes \citep{Yaglom}.  This interest can partially be explained by the mathematical well-posedness  of the problem and by the emergence of numerical methods combined with powerful computers. Conversely, the ``fuzzy''  problem involving inlet disturbances, pipe imperfections, and pipe roughness  has not attracted as much attention ~\citep{Pausch_2015}. In this article, we investigate both experimentally and theoretically   transition to turbulence in imperfect  channels.  In other words, we are not interested in reducing roughness or removing wall and inlet imperfections. Specifically, we strive to quantify  growing perturbations near the wall by their spectra and statistical properties, including probability densities and low- and high-order moments. Remarkably, we  observe that, while low-order moments are relatively independent of the experimental details, high-order moments are extremely sensitive to these details and may differ by orders of magnitudes. To describe our experimental observations, we propose a modified  Landau theory in which  all  imperfection-induced flow disturbances  are  treated as  added high-frequency noise. This theory agrees well with our experimental observations. An important consequence of this theory is that the noise  regularizes the Landau singularity of the relaxation time. It must be emphasized that, in this regime (i.e., in imperfect channels),  turbulence is constantly driven by  noise, which makes the problem very  different from the initial value problem  considered by \citet{Hof_2006}.

\section{Experiments}

\subsection{Experimental Set-up}

\begin{figure}
\centerline{\includegraphics[width=5.3 in]{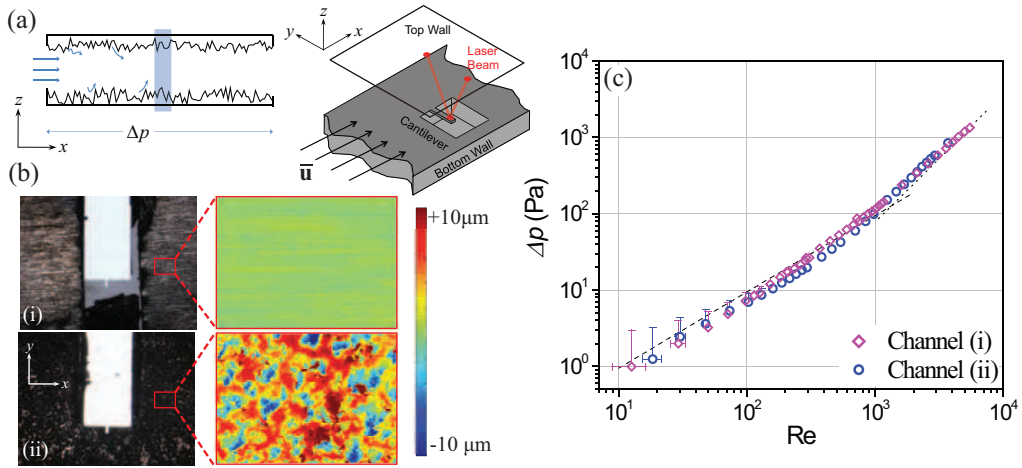}}
\caption{(a) Rectangular channel with linear dimensions $L_x \times L_y \times L_z = 35 \times 8 \times 0.9~\rm{mm}^3$. The test section is highlighted. The microcantilever chip is integrated to the middle of the bottom wall.  (b) Optical microscope (left) and scanning white light  interferometry images of the bottom walls of the two channels. The optical images are $6.2 \times 4.6~\rm{mm}^2$, and the interferometry images are $0.7 \times 0.5 ~\rm{mm}^2$. The color scale bar is the same for both images. (c) Double logarithmic plot of pressure drop $\Delta p$ as a function of Reynolds number in the channels.  Lines are fits to laminar (dashed) and Blasius (dotted) flow models with the transition around $\rm Re \sim 2\times 10^3$. Error bars are shown only when larger than symbols.}
\label{fig:schematic}
\end{figure}

We have performed our experiments in two nominally identical rectangular channels with linear dimensions  $L_x \times L_y \times L_z = 35 \times 8 \times 0.9 ~\rm{mm}^3$ and with hydraulic diameters   ${D}  \approx 1.6$ mm. A pressurized air source is used to establish air  flow in each channel.  The flow rate is monitored using a rotameter, and  the pressure drop $\Delta p$ between the inlet and the outlet is measured using a differential gauge. The near-wall flow in each channel is probed by a rectangular microcantilever with linear dimensions $l_x \times l_y \times l_z = 27.5  \times 225 \times 3 ~\rm{\mu m}^3$ [Fig.~\ref{fig:schematic}(a)-(b)]. In the first channel [henceforth channel (i)], the chip carrying the microcantilever is embedded in the bottom wall of the micro-channel by a surrounding aluminium structure that is cut from a $\sim 300-\rm{\mu m}$-thick smooth sheet (matching the height of the chip) and glued to the bottom wall [Fig.~\ref{fig:schematic}(b)]. The root mean square (r.m.s.) roughness on the wall is $\sim 300~\rm nm$.   In the second [henceforth channel (ii)], a groove, which matches the size of the cantilever chip closely,  is machined on the bottom channel wall  [Fig.~\ref{fig:schematic}(b)]. The r.m.s. roughness on this wall is significantly larger, $\sim 3~\rm \mu m$. In both channels, the test section is approximately 17 mm ($\approx 11D$) from the inlet. Because turbulence is driven and sustained by noise due to imperfections in our channels, the distance of the test section to the inlet is sufficient for the observation of the transition ~\citep{Zagarola_1998}.  Both channels have  smooth and transparent top walls, and  the motion of the tips of the microcantilevers  is read out using the optical beam deflection technique~\citep{Meyer_1988}. We confirm that our optical transducer remains linear in the explored parameter space.

We emphasize that the two channels, while  nearly identical in all macroscopic aspects (e.g., linear dimensions, inlet pipes), possess  different sets of microscopic imperfections on their walls and test sections, as seen in Fig.~\ref{fig:schematic}(b). Here, we are not concerned with the particulars of these imperfections, which may include but are not limited to surface roughness, wall asperities, bumps, edges  and so on. As we show below, even though these imperfections may lead to different random flows,  the lowest order statistics remain qualitatively similar.

%%*************************Fig1********************************************
%%*************************Fig1********************************************
%**************************************************************************
%**************************************************************************

\subsection{Pressure Drop}

The pressure drop $\Delta p$ between the inlets and outlets of both channels are shown in Fig.~\ref{fig:schematic}(c) as a function of the Reynolds number.   In the low Reynolds number range $0 \leq {\rm Re} \lesssim 10^3$, $\Delta p$ can be fit to that in a  Poiseuille flow using the nominal channel parameters with no free parameters. This is the dashed asympote  in Fig.~\ref{fig:schematic}(c). In the high $ \rm Re$ regime,   the data  appears  to converge to another asymptote (dotted line).  This  is  the pressure drop  calculated from the Colebrook equation in a rectangular duct~\citep{Jones_1976}, using only  experimentally available parameters. The pressure drops in both channels match closely.

%%*************************Fig2********************************************
%%*************************Fig2********************************************
\begin{figure}
%figure is 4.85 in wide
\centerline{\includegraphics[width=4.95 in]{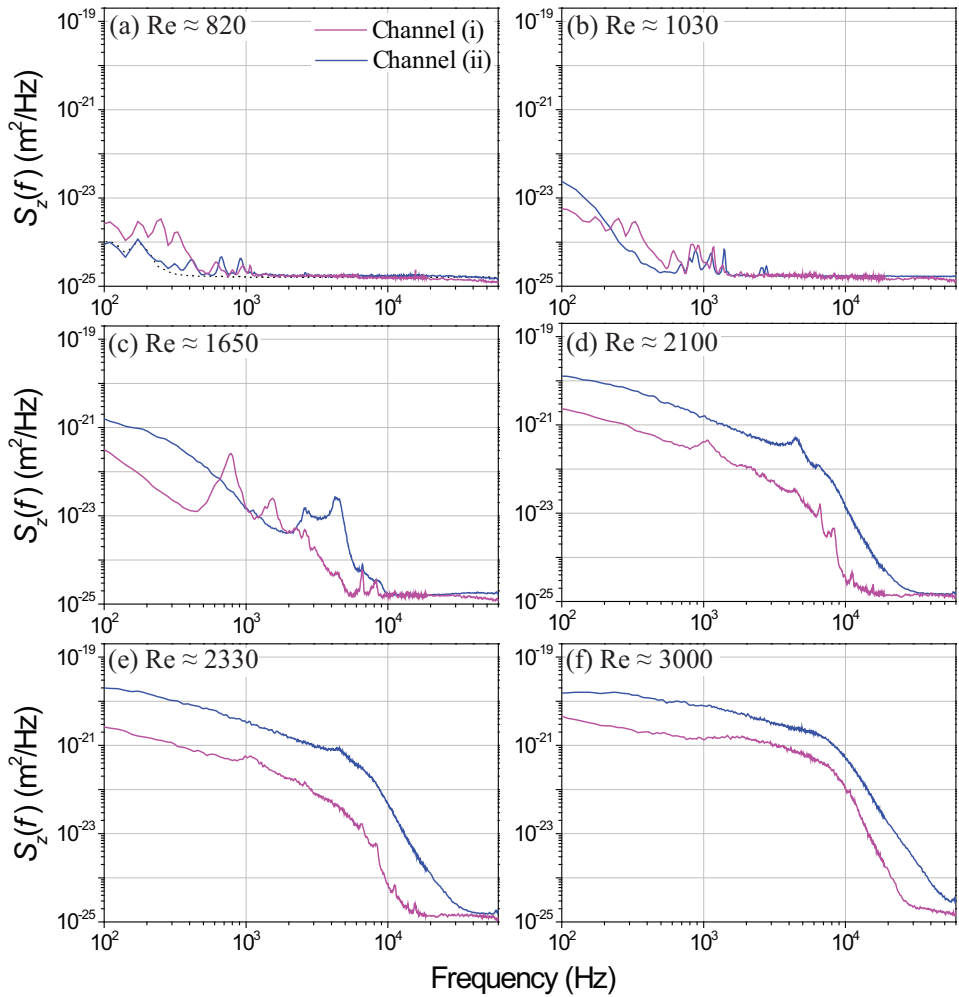}}
\caption{Power spectral density $S_z(f)$ of cantilever displacements for different Reynolds numbers for channels (i) and (ii). The dotted line in (a) shows the noise floor of the measurement; the noise floor and the spectra are indistinguishable at $\rm Re \approx 820$  for the frequency range $f \ge 300$ Hz. }
\label{fig:spectra}
\end{figure}
%**************************************************************************
%**************************************************************************

\subsection{Spectral Properties}

We first turn to the spectral properties of near-wall fluctuations. Flow forces  act on the microcantilever and give rise to mechanical fluctuations. The microcantilever has a sharply-peaked resonance at $100$ kHz with a linewidth of $500$ Hz. Its linear response function is frequency-independent in the frequency range $f < 80$ kHz and can be approximated as $|G(f)| \approx 1/\kappa$, with $\kappa$ being the cantilever  spring constant and $\kappa\approx 3$ N/m. Thus, the linear relation, ${S_z(f)} \approx {{ {S_F(f)}}/\kappa^2}$, between the spectral densities of the force ${S_F(f)}$ and the cantilever displacement ${S_z(f)}$ remains valid below $f < 80$ kHz. At very low flow rates (${\rm Re}\lesssim 800$),  ${S_z(f)}$ becomes obscured by measurement noise  because the flow cannot generate detectable cantilever motion.

The spectral densities of the cantilever fluctuations $S_z(f)$ obtained for the two channels are shown at different  Reynolds numbers in Fig.~\ref{fig:spectra}. For ${\rm Re} \lesssim 1000$ [Fig.~\ref{fig:spectra}(a) and (b)], the spectra for both cases are barely above the noise floor (dotted black line) and appear  similar for the most part. For ${\rm Re} \ge 1600$ [Fig.~\ref{fig:spectra}(c)-(f)],  small differences between the two cases can be noticed.  First, the fluctuations in channel (ii) with the rougher wall are  larger than those in channel (i) by an order of magnitude; second, the spectrum in channel (ii)  extends to slightly higher frequencies.   As Reynolds number is increased,  both data traces increase monotonically and smoothly. Interestingly,  $S_z(f) \propto 1/f$ for both channels in the low frequency range.

%
%%*************************Fig4********************************************
%%*************************Fig4********************************************
\begin{figure}
%Figure is 4.88 in wide
\centerline{\includegraphics[width=5.09 in]{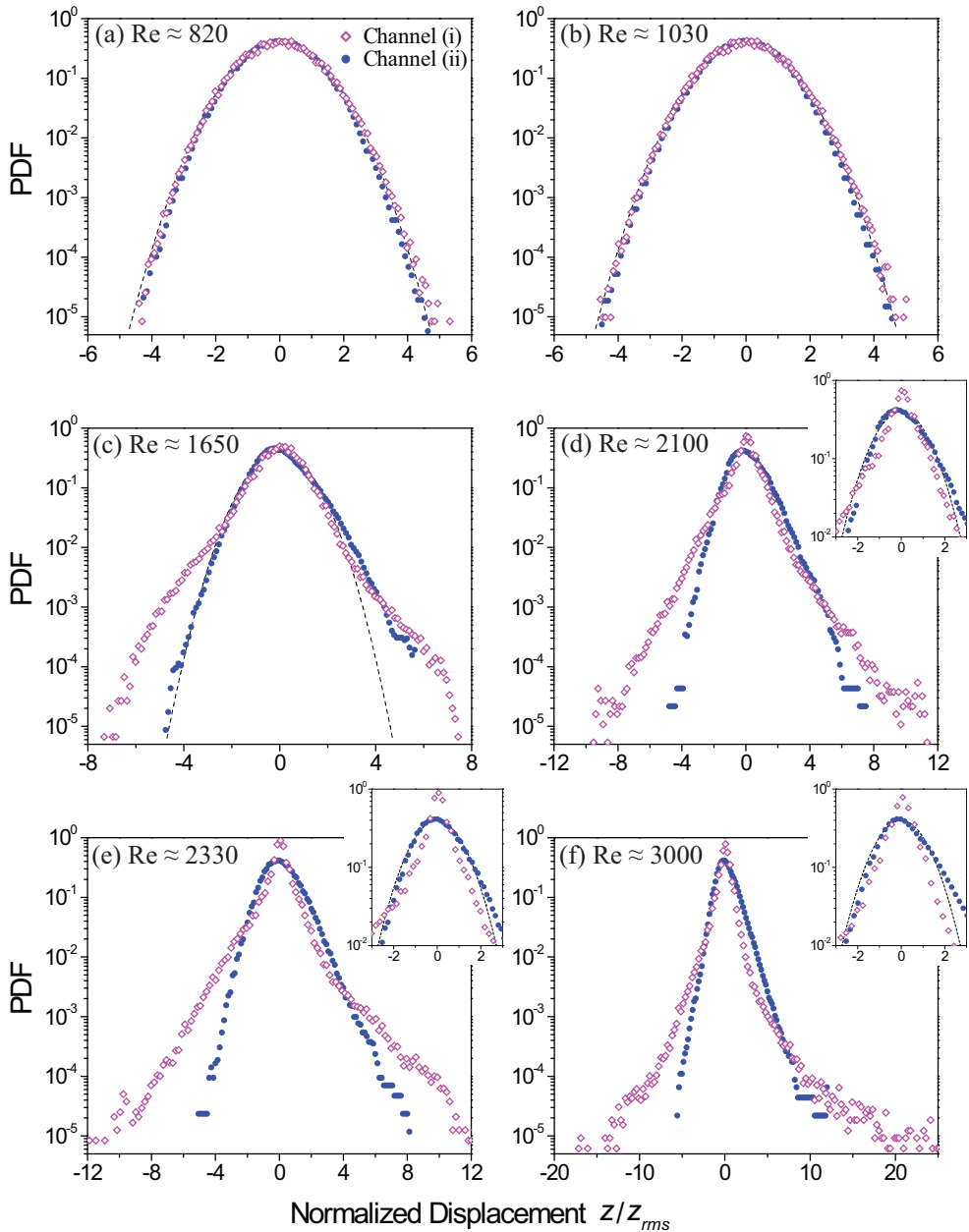}}
\caption{PDFs at different Reynolds numbers for channels (i) and (ii). The $x$-axes are in units of $z_{rms}$,  which are different for each data set (see table~\ref{table:moments}). Dotted lines are Gaussians. }
\label{fig:pdfs}
\end{figure}
%**************************************************************************

\subsection{Statistical Properties}

Next, we turn to the statistical properties of the near-wall fluctuations. We collect  long-time traces of the cantilever amplitudes filtered  in the frequency range $100~{\rm Hz} < f <  30~{\rm kHz}$ to remove the high-frequency resonant oscillations. We then sample the time data every $3~\mu \rm s$, obtaining  $\sim 10^6$ data points. Because we are interested in single-point probability density functions (PDFs), we do not worry about over-sampling in comparison to other relaxation times in the measurement, e.g., the ring-down time of the cantilever.

\begin{figure}
\centerline{\includegraphics[width=5.21 in]{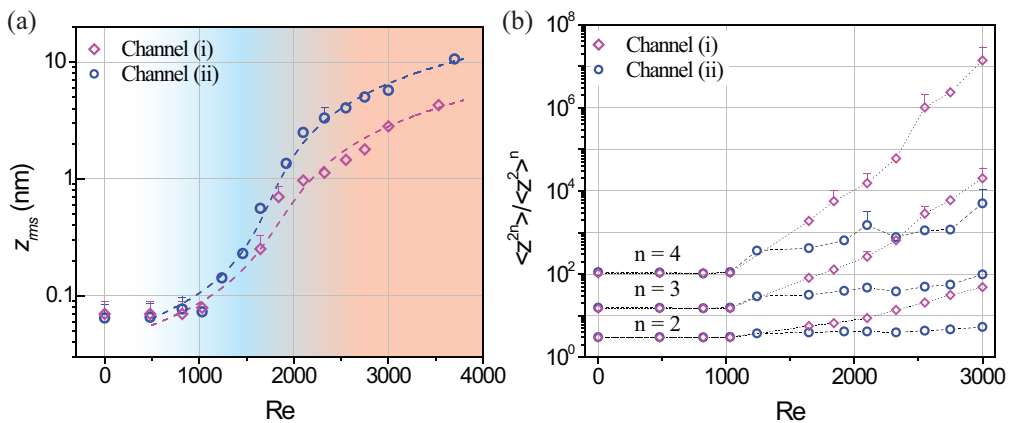}}
\caption{(a) Cantilever r.m.s.  displacements, $z_{rms}$, versus Re for channels (i) and (ii). Dashed lines are fits to the results of the noisy Landau Equation, Eq.~(\ref{eq:result_noisy}), with  parameters given in Table~\ref{table:landau}. (b) Normalized high-order moments. Error bars are estimated from convergence tests performed by computing the moments for a subset of the data  (e.g., half). When not shown,  error bars are smaller than symbol size.}
\label{fig:variance}
\end{figure}

The  PDFs  for different Reynolds numbers are plotted  in Fig.~\ref{fig:pdfs}. For $\rm Re \lesssim 1000$, our measurements are dominated by technical noise, and the PDFs in both channels (i) and (ii) are perfectly Gaussian [Fig.~\ref{fig:pdfs}(a) and (b)].  In the range ${\rm Re} \ge 1600 $, both PDFs go through dramatic changes  as seen in Fig.~\ref{fig:pdfs}(c)-(f). First, one notices a substantial broadening of the tails of the PDF in channel (i), indicating the presence of  strong wall velocity/pressure bursts, which may reach  the bulk \citep{Yakhot_2010}. The PDFs observed in channel (ii) for $\rm Re \ge 1600$ suggest that the flow here is somewhat more homogenous [Fig.~\ref{fig:pdfs}(c)-(f)] compared to that in channel (i), but wall bursts make the PDFs more asymmetric and intermittent [Fig.~\ref{fig:pdfs}(e)]. This trend continues with increasing Reynolds number [Fig.~\ref{fig:pdfs}(f)]. The dotted lines in the insets show that the PDFs in channel (i) cannot be fit to Gaussians, even at very small displacements; an exponential decay seems perhaps more appropriate. The differences between the two flows  can also be clearly seen in the moments of the PDFs listed in Table~\ref{table:moments}. The moments are also plotted in Fig.~\ref{fig:variance}, with  ${z_{rms}}={\left\langle {{z^2}} \right\rangle }^{1/2}$ (brackets indicate averaging) in Fig.~\ref{fig:variance} (a) and normalized high-order moments in  Fig.~\ref{fig:variance} (b). The high-order moments obtained in the two channels differ by orders of magnitude. These observations can be summarized as follows. The random flow in channel (i) is more intermittent compared to that in channel (ii); however, the intensity of fluctuations in channel (ii) is larger. Based upon these observations, it may not be incorrect to conclude that these are two very different random flows. While the exact source of the differences in these two flows is impossible to identify, it must be due to the different set of imperfections in the two channels.

We return to the r.m.s displacement of the cantilever, ${z_{rms}}={\left\langle {{z^2}} \right\rangle }^{1/2}$,  as a function of Reynolds number, shown in Fig.~\ref{fig:variance} (a). The dashed lines are fits to our proposed equation, discussed below.  We note that ${z_{rms}}$ values obtained by integrating the spectra  and from the time domain measurements agree closely.  There are three regions in the plot marked with different shadings. The first region, ${\rm{Re}} \lesssim 800$, is dominated by technical noise and does not provide any insight. In the second region (blue) where $\rm 1000 \lesssim Re \lesssim 1600$,  the magnitude of the cantilever fluctuations are of the same order for both channels (i) and (ii), ${z_{rms}} \sim 1$ nm. In the third region (pink), the observed r.m.s. amplitude of near-wall fluctuations in the rough channel (ii) are  larger than those in the smooth channel (i). The slope changes for both data traces around $\rm Re \approx 2000$, suggesting the onset of more sustained perturbations. The data traces appear to increase parallel to each other for $\rm Re \gtrsim 2000$.

\begin{table}
\begin{center}
\begin{tabular}{l|c|c|c|c|c|c|c|c}
Re  & \multicolumn{2}{c|}
{$ {\left\langle {{z^2}} \right\rangle }^{1/2}  $ (nm)} &
  \multicolumn{2}{c|}
{${\frac{\langle z^4 \rangle}{\langle z^2 \rangle ^2}}$} &
\multicolumn{2}{c|}
{${\frac{\langle z^6 \rangle}{\langle z^2 \rangle ^3}}$} &
\multicolumn{2}{c}
{${\frac{\langle z^8 \rangle}{\langle z^2 \rangle ^4}}$}\\
 & \multicolumn{1}{c}{(i)} & \multicolumn{1}{c|}{(ii)} & \multicolumn{1}{c}{(i)} & \multicolumn{1}{c|}{(ii)} & \multicolumn{1}{c}{(i)} & \multicolumn{1}{c|}{(ii)} & \multicolumn{1}{c}{(i)} & \multicolumn{1}{c}{(ii)}\\
\hline
0    & 0.07 & 0.06 & 2.99 & 3.04 & 14.9 & 15.5 & 104  & 111  \\
480  & 0.07 & 0.07 & 3.00 & 3.04 & 15.0 & 15.4 & 105  & 109  \\
820  & 0.07 & 0.08 & 2.99 & 3.01 & 15.0 & 15.0 & 106  & 103  \\
1030 & 0.08 & 0.07 & 3.00 & 3.06 & 15.1 & 15.6 & 107  & 111  \\
1240 &   -  & 0.14 &   -  & 3.75 &  -   & 29.2 &  -   & 366  \\
1650 & 0.25 & 0.56 & 5.61 & 3.92 & 81.5 & 31.8 & 1890 & 421  \\
2100 & 0.97 & 2.50 & 8.75 & 4.15 & 261  & 47.3 & 15400& 1500 \\
2330 & 1.13 & 3.33 & 13.7 & 3.94 & 654  & 38.1 & 60400& 768  \\
2550 & 1.46 & 4.05 & 20.7 & 4.34 & 2840 & 49.6 & $1.02 \times 10^6$ & 1120 \\
2750 & 1.79 & 5.02 & 31.2 & 4.65 & 5960 & 55.4 & $2.36 \times 10^6$ & 1170 \\
3000 & 2.82 & 5.73 & 48.4 & 5.33 & 20100& 97.6 & $1.39 \times 10^7$ & 5010 \\
\end{tabular}
\caption{Moments.}{\label{table:moments}}
\end{center}
\end{table}
%**************************************************************************
%**************************************************************************
%%*************************Fig4********************************************
%%*************************Fig4********************************************
%**************************************************************************
%**************************************************************************

\section{Theory}

The similarities in the two data traces in Fig.~\ref{fig:variance}(a) suggest that there may be  common underlying physics  to  both cases. In order to describe both flows by a single equation, we return to the Landau theory discussed above.  We  incorporate the non-idealities into the Landau theory by  considering a general additive noise term $\phi(t)$. We assume that  a  high-frequency random Gaussian force  $\phi(t)$  defined by the correlation function $\overline{\phi(t)\phi(t')} = 2\overline{ \phi^{2}}\delta(t-t')$  {stirs} the fluid. Then, Eq.~(\ref{eq:NS_perturbation}) is modified as
\begin{equation}
\frac{\partial {{\bf v}}}{\partial t}+{\bf \bar u \cdot \nabla \bf v}+{\bf v \cdot \nabla \bar u}+ {\bf v \cdot \nabla v}=-{\frac{{\bf \nabla} p_1} {\rho}}+\nu\nabla^{2}{\bf v} +\phi(t).
\label{eq:NS_pert2}
\end{equation}
Repeating Landau's arguments leading to Eq.~(\ref{eq:landau_nopert}), we write
\begin{equation}
\frac{\partial A}{\partial t} = \gamma_{1} A - \frac{\alpha}{2} A^{3}+\phi(t).
\label{eq:noisy_Landau_1}
\end{equation}
Averaging over high-frequency fluctuations gives a modified Landau equation for the slow mode
\begin{equation}
\frac{d|A|^{2}}{dt} = 2{\gamma _1}{|A{|^2}} - \alpha {|A{|^4}} + \overline {{\phi ^2}}.
\label{eq:noisy_Landau}
\end{equation}
In a force-driven flow with initial condition $A_{0}=0$, the solution to Eq.~(\ref{eq:noisy_Landau}) is
 \begin{equation}
|{A(t)}|^2  =  \frac{\gamma_1}{\alpha} + \frac{\beta_- - \beta_+ e^{-\frac{t}{\tau}}}{\beta_- + \beta_+ e^{-\frac{t}{\tau}}} \left( \beta_+ - \frac{\gamma_1}{\alpha} \right),
\end{equation}
where $\beta_{\pm} = \pm\frac{\gamma_1}{\alpha} + \sqrt{\frac{{\gamma_1}^2}{{\alpha}^2} + \frac{\overline{\phi^2}}{\alpha}}>0$  with $\tau=\frac{1}{2\sqrt{\gamma^{2}_{1}+\overline{\phi^{2}}{\alpha}}}$.
This gives in the long-time limit
\begin{equation}
|{A(\infty)}|^2  = \beta_{+}= \frac{\gamma_1}{\alpha} + \sqrt{\frac{{\gamma_1}^2}{{\alpha}^2} + \frac{\overline{\phi^2}}{\alpha}}.
\label{eq:result_noisy}
\end{equation}
Remembering that in Landau theory $\gamma_1=c({\rm Re- Re_{cr}})$, we notice an important consequence of noise: the relaxation time $\tau$ remains finite in the limit $\rm Re\rightarrow Re_{cr}$, in contrast to the ``critical slowing down" discussed above.  Thus,  the external noise source  regularizes  the dynamics around the transition point. In addition, Eq.~(\ref{eq:noisy_Landau_1}) indicates  that,  when   $A$ is small in the low $\rm Re$  limit so that the ${\cal O}(A^{3})$  contributions can be neglected,  $A$ obeys Gaussian statistics.

%%*************************Table*******************************************
%%*************************Table*******************************************
\begin{table}
\begin{center}
\begin{tabular}{lccc}
   &  $\rm Re_{cr}$ & $\gamma_1 / \alpha~{\rm(m^2/s^2)}$ & $\overline{\phi^{2}} / \alpha~{\rm(m^4/s^4)}$ \\
\hline
%Case (i)    & $1850$ & $9.1 \times 10^{-4} \times {\rm(Re-Re_{cr})}$ & $0.11$   \\
%Case (ii)   & $1700$ & $1.3 \times 10^{-2} \times {\rm(Re-Re_{cr})}$ & $1.46$   \\
Channel (i)    & $1850$ & $0.14 \times {\rm(Re-Re_{cr})}$ & $2.37 \times 10^{3}$   \\
Channel (ii)   & $1750$ & $0.29 \times {\rm(Re-Re_{cr})}$ & $5.41 \times 10^{3}$   \\
\end{tabular}
\caption{Fit parameters used [see Eq.~(\ref{eq:result_noisy})]. }{\label{table:landau}}
\end{center}
\end{table}
%**************************************************************************
%**************************************************************************

The obtained results can be used to quantitatively explain the experimental data of Fig.~\ref{fig:variance}(a). The time-dependent force acting on the cantilever is
\begin{equation}
{\bf F} = F{\bf \hat z}=-\int_S p_1 \cdot {\bf n} dS \approx \frac{\partial p_1}{\partial z} V {\bf \hat z},
\end{equation}
where ${\bf \hat z}$ is the unit vector, and $V$ is the volume of the cantilever. We simplify the problem by approximating our channel as  a  long and wide rectangular (planar) duct and, neglecting noise,  write the perturbation equation in the $z$ direction from Eq.~(\ref{eq:NS_pert2})  as
\begin{equation}
\frac{1}{\rho}\frac{\partial p_1}{\partial z} = - \frac{\partial v_z}{\partial t} - {\frac{3}{2}}U\left( 1 - { \frac{4z^2}{H ^2}} \right) \frac{\partial v_z}{\partial x} +
\nu \nabla^2 v_z +{\cal O}(\frac{\partial{v^{2}}}{\partial z}).
\label{eq:near_wall_channel}
\end{equation}
In close proximity to the wall,  $z\approx H/2$, and the second term on the right hand side of Eq.~(\ref{eq:near_wall_channel}) is small. Numerical simulations \citep{Lee}, where statistics of acceleration in close proximity to a wall was studied, suggest that bursts are dominated by the $z$-component of the velocity ($v^{2}={\cal O}({v_{z}}^{2})$) and the viscous term is unimportant.   Therefore,
\begin{equation}
\frac{1}{\rho}\frac{\partial p_1}{\partial z} \approx - \frac{\partial v_z}{\partial t} +{\cal O}(\frac{\partial{{v_{z}}^{2}}}{\partial z}).
\label{eq:result_2}
\end{equation}
Here,  $\partial_{t}v_{z}={\cal O}(v_{z}U/H)$. To find an order of magnitude estimate for $\partial_{z}({{v_{z}}^2})$,  we  extrapolate the   results of \citet{Yakhot_2010}, which shows that,  for $\rm Re<10^{5}$, the velocity of wall bursts are $v_{z}={\cal O}(U/10)$. This means that the magnitude of the two terms on the right hand side of Eq.~(\ref{eq:result_2}) are comparable, and ${\cal O}({v_{z}}^{2})$ must be accounted for. Thus, we deduce  $\partial_{z} p_{1}={\cal O}({v_{z}}^{2}/H)$ and write
\begin{equation}
z_{rms} \sim {\rho V {v_z}^2}/{\kappa H} \propto |{A(\infty)}|^2.
\end{equation}
This exercise provides the fits (dashed lines) shown in Fig.~\ref{fig:variance} (a) with the parameters in Table~\ref{table:landau}. The ultimate justification for the above arguments comes from the agreement between experiment and theory in Fig.~\ref{fig:variance} (a). The following simple order of magnitude estimate further bolsters our confidence. Assuming  ${v_z}^2/U^2\approx  0.008$ around $\rm Re_{cr}$ ~\citep{Yakhot_2010} and using the  experimentally available parameters ($V\approx 2\times 10^{-14}$ m$^3$, $H\approx 10^{-3}$ m, $\kappa \approx 3$ N/m), we find $z_{rms} \approx 10^{-10}$ m, close to the experimental values.

\section{Summary and Outlook}

In transitional flows in imperfect channels, high-order moments of near-wall fluctuations appear to be extremely sensitive to the specifics  of the imperfections in the channels. In the two nominally identical channels with different sets of imperfections presented here,  the high-order moments of the near-wall fluctuations  differ by orders of magnitude.  On the other hand, low-order moments  remain relatively independent of the details of the imperfections. In our experiments,  $z_{rms}$ values seem to depend directly on the wall roughness. Additionally, $z_{rms}$ in these  flows can be accurately described by the noisy Landau equation presented. The applicability of the noisy Landau equation to these different flows is probably due to the fact that the  noise term is taken to be at high frequencies as compared to the slow mode in question. In other words, the stirring of the fluid occurs at high frequencies. This assumption is eminently reasonable because microscopic surface asperities giving rise to roughness are at high spatial frequency compared to any length scales of the flow. Our preliminary results suggest that inlet disturbances may also be accounted for by the noisy Landau equation. Along the same lines, we believe that even  thermal fluctuations in the fluid would likely result in a similar regularization of the Landau equation. Finally, the phenomena observed here may have important consequences for heat and mass transfer in wall-bounded flows;  these  will be discussed in detail in future work.

%*******************Thanks************************************************************
%*************************************************************************************

We acknowledge support from the US NSF through grant Nos. CMMI-0970071 and DGE-1247312. We are grateful to T. Kouh for intial measurements, and K. R. Sreenivasan and J. Schumacher for valuable comments.

%*************************************************************************************
%*************************************************************************************

%\section{Citations and references}
%All papers included in the References section must be cited in the article, and vice versa. Citations should be included as, for example ``It has been shown \citep{Rogallo81} that...'' (using the {\verb}\citep} command, part of the natbib package) ``recent work by \citet{Dennis85}...'' (using {\verb}\citet}).
%The natbib package can be used to generate citation variations, as shown below.\\
%\verb#\citet[pp. 2-4]{Hwang70}#:\\
%\citet[pp. 2-4]{Hwang70} \\
%\verb#\citep[p. 6]{Worster92}#:\\
%\citep[p. 6]{Worster92}\\
%\verb#\citep[see][]{Koch83, Lee71, Linton92}#:\\
%\citep[see][]{Koch83, Lee71, Linton92}\\
%\verb#\citep[see][p. 18]{Martin80}#:\\
%\citep[see][p. 18]{Martin80}\\
%\verb#\citep{Brownell04,Brownell07,Ursell50,Wijngaarden68,Miller91}#:\\
%\citep{Brownell04,Brownell07,Ursell50,Wijngaarden68,Miller91}\\
%The References section can either be built from individual \verb#\bibitem# commands, or can be built using BibTex. The BibTex files used to generate the references in this document can be found in the zip file at http://journals.cambridge.org/\linebreak[3]data/\linebreak[3]relatedlink/\linebreak[3]jfm-ifc.zip.\\

%\bibliography{jfm-references}
%\bibliographystyle{jfm}

\end{document}